\documentclass{jpp}
\usepackage{graphicx}
\usepackage[utf8]{inputenc}
\usepackage[T1]{fontenc}
\usepackage{amsmath}
\usepackage{xcolor}
\usepackage{changes}
\definechangesauthor[color=olive]{AS}
\definechangesauthor[color=violet]{AV}
\definechangesauthor[color=violet]{SL}
\definechangesauthor[color=blue]{LC}

\shorttitle{Coarse-grained dynamics of collisionless plasma}
\shortauthor{L. Barbieri et al.}
\title{
Extended temporal coarse-graining in a stratified and confined plasma under thermal fluctuations
}
\author{Luca Barbieri\aff{1}
  \corresp{\email{luca.barbieri@obspm.fr}},
  Simone Landi\aff{2,3},
  Lapo Casetti\aff{2,3,4},
  Andrea Verdini\aff{2,3}
 }
\affiliation{\aff{1} LIRA, Observatoire de Paris, Université PSL, Sorbonne Université, Université Paris Cité, CY Cergy Paris Université, CNRS, 92190
Meudon, France
\aff{2}Dipartimento di Fisica e Astronomia, Universit\`a di Firenze, via Giovanni Sansone 1, Sesto Fiorentino, I-50019, Italy
\aff{3}INAF - Osservatorio Astrofisico di Arcetri, Largo Enrico Fermi 5, Firenze, I-50125, Italy
\aff{4} INFN - Sezione di Firenze, via Giovanni Sansone 1, Sesto Fiorentino, I-50019, Italy
}
\begin{document}
\maketitle

\begin{abstract}
We present an extended investigation of a recently introduced model of gravitationally confined, collisionless plasma (\citealt{barbieri2023temperature}), which showed that rapid temperature fluctuations at the base of the plasma, occurring on timescales much shorter than the electron crossing time, can drive the system into a non-thermal state characterized by anti-correlated temperature and density profiles, commonly referred to as temperature inversion. To describe this phenomenon, a temporal coarse-graining formalism was developed (\citealt{Barbieri2024b}). In this work, we generalize that approach to cover regimes where the timescales of temperature fluctuations are comparable to or exceed the electron crossing time. We derive a set of kinetic equations that incorporate an additional term arising from the coarse-graining procedure, which was not present in the earlier formulation. Through numerical simulations, we analyze the plasma dynamics under these broader conditions, showing that the electric field influences the system when fluctuation timescales approach the electron crossing time. However, for timescales much larger than the proton crossing time, the electric field becomes negligible. The observed behaviours are interpreted within the framework of the extended temporal coarse-graining theory, and we identify the regimes and conditions in which temperature inversion persists.

\end{abstract}
\keywords{plasma dynamics, plasma properties, plasma nonlinear phenomena}

\maketitle
\section{Introduction}

\indent 
The lower solar atmosphere consists of a dense, collisional plasma in thermal equilibrium at temperatures around 10000K. In contrast, the outer atmospheric layer, the solar corona, is composed of a much more rarefied but significantly hotter plasma, with temperatures reaching 1–2 million K. Understanding the physical mechanisms responsible for this dramatic temperature increase remains one of the central unresolved challenges in solar and plasma physics, commonly referred to as the coronal heating problem (\citealt{Klimchuk_2006,2012coronalheating,2005ApJ...618.1020G,Rappazzo:2008vl,2013ApJ...773L...2R,2015RSPTA.37340265W,Heyvaerts_Priest_1983,Ionson_1978,2020A&A...636A..40H,Pontieu:2011vg, Scudder1992a,Scudder1992b,Hau_2025}).

A recent study introduced a novel kinetic $N$ particle model of coronal loops in the solar atmosphere (\citealt{barbieri2023temperature}), showing through numerical simulations that rapid, stochastic temperature increments in the high chromosphere, faster than the electron crossing time of the loop, can drive the plasma into a non-equilibrium stationary state characterized by anti-correlated temperature and density profiles, a phenomenon referred to as temperature inversion. In a subsequent work (\citealt{Barbieri2024b}), the model was analyzed within a kinetic framework, leading to the development of a temporal coarse-graining formalism capable of describing the long-time behaviour of the system. It was proven that the plasma dynamics can be captured by a pair of coupled Vlasov equations for coarse-grained distribution functions, with stationary solutions exhibiting suprathermal velocity distributions. These suprathermal tails are self-consistently produced by the temperature increments, and the temperature inversion naturally arises as a result of gravitational filtering (i.e., more energetic particles in the suprathermal tails can rise higher than the others in the gravity well (\citealt{Scudder1992a,Scudder1992b})). Remarkably no fine-tuning of parameters is required to produce temperature inversion, as long as the characteristic timescale of thermal fluctuations remains shorter than the electron crossing time. Moreover, these results are independent of the self-consistent electrostatic field. The formalism was subsequently applied to solar-type main-sequence stars, predicting the presence of a hot corona for all such stars \citep{Barbieri2025}.

In the present study, we extend this analysis by relaxing the restriction on fluctuation timescales and generalizing the temporal coarse-graining approach accordingly. We show that the dynamics can still be described by two coupled kinetic equations, but with an additional term arising from the coarse-graining procedure itself. This term becomes relevant when temperature fluctuations occur on timescales comparable to the electron crossing time, resulting in species-dependent density and temperature profiles and a suppression of gravitationally driven temperature inversion.If the fluctuation timescales become much larger than the proton crossing time, the system exhibits oscillatory behaviour between multiple thermal configurations. In this regime, the stationary coarse-grained distribution corresponds to a superposition of thermal states, and temperature inversion can still be recovered under specific definitions of the coarse-grained temperature.

The paper is organized as follows. In Section \ref{sec2}, we review the kinetic $N$-particle model introduced in (\citealt{barbieri2023temperature}).In Section \ref{sec3}, we develop the extended temporal coarse-graining formalism and show that it recovers the earlier formulation as a limiting case under appropriate assumptions. Section \ref{sec4} contains a detailed numerical analysis of plasma dynamics across different fluctuation timescales, interpreted within the new theoretical framework. Finally, Section 5 summarizes the main results and outlines potential future developments.

\section{The two-component gravitationally bound plasma model}\label{sec2}

\begin{figure}
    \centering    \includegraphics[width=0.99\columnwidth]{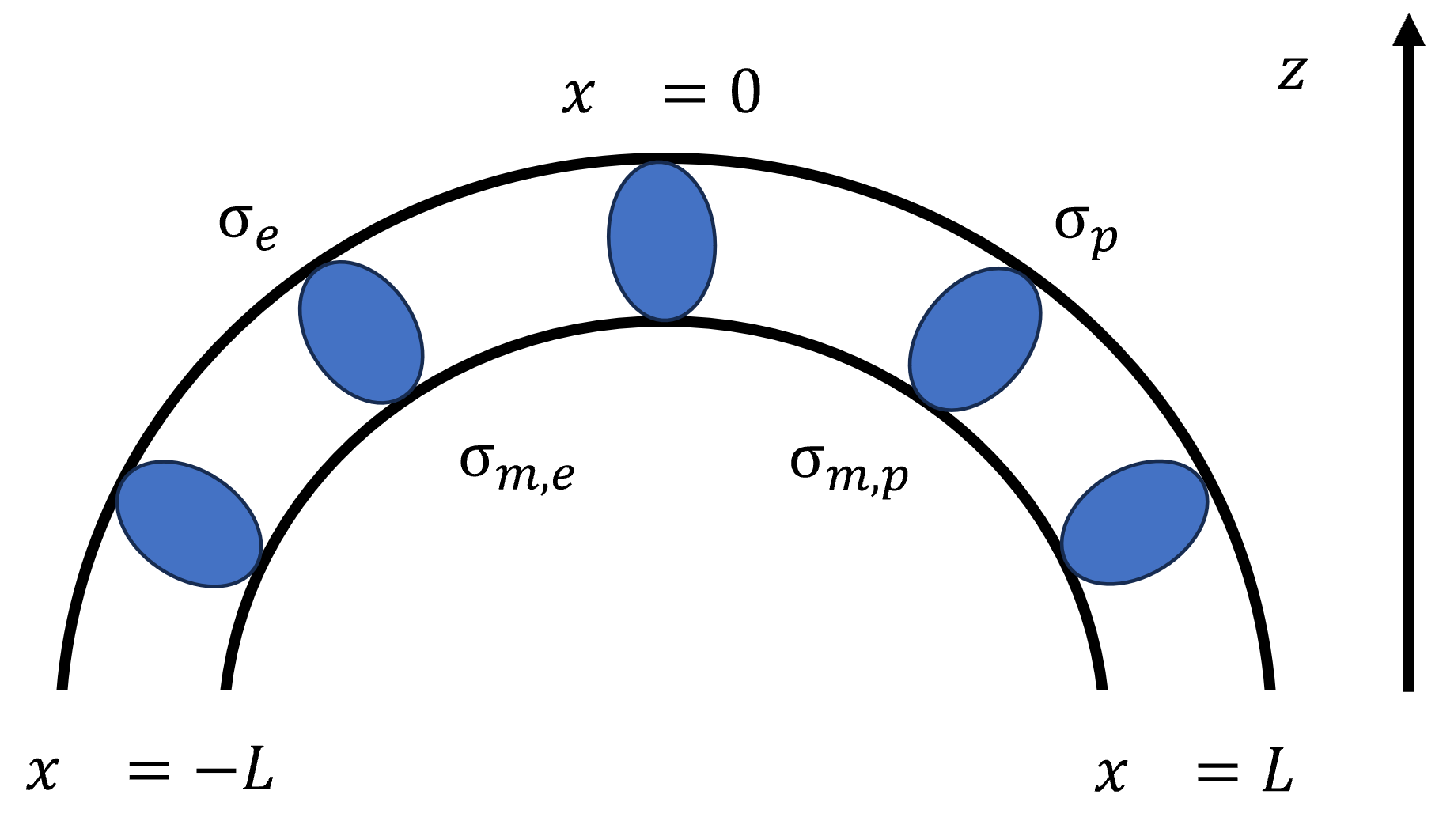}
    \caption{Schematic diagram of the two-component plasma loop model. The vertical axis, designated as $z$, represents the altitude within the atmosphere, while the curvilinear abscissa is denoted as $x$. The symbols $\sigma_{m,\alpha}= m_{\alpha} n_S$ and $\sigma_{\alpha}=e_{\alpha} n_S$, with $\alpha=\{e,i\}$ are the surface mass density and the surface charge density of the species $\alpha$ while $n_S$ is the surface number density.
    }
    \label{fig:Fig1}
\end{figure}

Let us now briefly describe the model for geometrically confined plasma structures, specifically the coronal loops that are ubiquitous in the Sun’s atmosphere (see, e.g., \citealt{2005psci.book.....A}) introduced by \citealt{barbieri2023temperature,Barbieri2024b}. The loop is modeled as a semicircular tube of length $2L$ and cross-sectional area $S$, with the charge distribution discretized into $n_S$ density sections. These consist of $2N$ sections, each containing electrons of charge $-e$ and mass $m_e$, and protons of charge $e$ and mass $m_i$ . All particles are subjected to an external field comprising the gravitational field and the Pannekoek-Rosseland electrostatic field (see \citealt{Pannekoek_1922,Rosseland_1924}). For an in-depth treatment see \citealt{belmont2013collisionless}. It is assumed that all quantities are symmetric with respect to the apex of the loop. Electrostatic interactions are treated using a multimode approximation known as the Hamiltonian Mean Field (HMF) model (\citealt{Elskens2019,AntoniRuffo:pre1995,Chavanis2005}), in which the electrostatic potential is expanded in Fourier series and truncated in the first mode, as discussed in \citealt{Barbieri2024b}. A scheme of the two-component loop plasma model is shown in Fig. \ref{fig:Fig1}. Under these assumptions, the equations of motion for each particle $j$ are given by
\begin{equation}\label{equationsofmotion}
m_{\alpha} \ddot{x}_{j,\alpha}=eE \left(x_{j,\alpha}\right)+g \frac{\sum_{\alpha \in \{e,p\}}m_{\alpha}}{2}\sin{\left(\frac{\pi x_{j,\alpha}}{2L}\right)}\,,
\end{equation}
where $e_{\alpha}$ is equal to $e$ for protons and $-e$ for electrons while the electric field $E$ reads as
\begin{equation}\label{electricfield}
E(x_{j,\alpha})=8~\mathrm{sign}{\left(e_{\alpha}\right)}e \cdot n_SN Q \sin{\left(\frac{\pi x_{j,\alpha}}{L}\right)}\,,
\end{equation}
the parameter $Q$ is given by
\begin{equation}\label{Chargeimbalances}
Q=\sum_{\alpha \in \{e,i\}}\mathrm{sign}{\left(e_{\alpha}\right)}q_{\alpha}\,,
\end{equation}
the quantities $q_{n,\alpha}$ are given by
\begin{equation}\label{Stratificationparameters}
    q_{\alpha}=\frac{1}{N}\sum_{j=1}^{N} \cos{\left(\frac{\pi x_{j,\alpha}}{L}\right)}\,, 
\end{equation}
$\alpha \in \{e,i\}$ denotes the species (here electrons or protons), $g={G M_{\odot}}/{R_{\odot}^2}$ is the
gravitational field at the surface of the Sun where $M_{\odot}$ is the solar mass and $R_{\odot}$ is the solar radius, and $x$ is the spatial coordinate (i.e. the curvilinear abscissa of the loop). Henceforth, the parameter $Q$ is referred to as the charge imbalance parameter, and the $q_{\alpha}$ as the stratification parameters. If the majority of particles of species $\alpha$ are symmetrically concentrated at the base of the loop, then $q_{\alpha} \approx -1$; a uniform distribution yields $q_{\alpha} \approx 0$, and a concentration at the apex results in $q_{\alpha} \approx 1$. A non-zero $Q$ therefore implies a net charge imbalance in the system.

The system is assumed to be in ideal thermal contact with a thermal boundary that mimics a fully collisional chromosphere. 


\subsection{The system of units}
The following units are defined for velocity, mass and length:
\begin{equation}\label{Setsofunits}
\begin{gathered}
    v_0=\sqrt{\frac{k_B T_{0}}{m_e}}, \quad m_0=m_e, \quad  L_0=\frac{L}{\pi} \,.
\end{gathered}
\end{equation}
Using these units, the equations of motion given in Eq.\ \eqref{equationsofmotion} can be rewritten in dimensionless form as
\begin{equation}\label{HMF2Sdimensionless}
        M_{\alpha} \ddot{\theta}_{j,\alpha}= \mathrm{sign}{\left(e_{\alpha}\right)} C E\left(\theta_{j,\alpha}\right)+\tilde{F}\left(\theta_{j,\alpha}\right) \quad,
\end{equation}
where the external and electrostatic forces are
\begin{equation}\label{HMF2Sdimensionless2}
        \tilde{F}\left(\theta_{j,\alpha}\right)= \tilde{g} \sin{\left(\frac{\theta_{j,\alpha}}{2}\right)}, \quad E \left(\theta \right)=  Q\sin{\left(\theta \right)} \,,
\end{equation}
with the parameters $Q$ and $q_{\alpha}$ given by
\begin{equation}\label{HMF2Sdimensionless3}
        Q=\sum_{\alpha \in \{p,e\}} \mathrm{sign}{\left(e_{\alpha} \right)}q_{\alpha},\quad q_{\alpha}=\frac{1}{N}\sum_{j=1}^{N}\cos{\left(\theta_{j,\alpha}\right)} \,.
\end{equation}
In these equations, $M_{\alpha}$ equals the mass ratio $M=m_i/m_e$ for protons, and $M=1$ for electrons. The coordinate $\theta$ represents the dimensionless spatial position. The dynamics of the system is thus fully characterized by the three dimensionless parameters
\begin{equation}
        M=\frac{m_i}{m_e}\,, \quad
        C=\frac{8 e^2 L^2 n_0}{\pi k_B T_0}\,, \quad
        \tilde{g}=\frac{g L \left(m_i+m_e \right)}{2 \pi k_B T_{0}}\,, 
\end{equation}
where $n_0=n_S N/L$ is the average density of a given species. The parameters $C$ and $\tilde{g}$ quantify the strengths of the electrostatic interaction and the external field, respectively, in units of thermal energy. Unless otherwise specified, all equations and plotted quantities hereafter are expressed in these dimensionless units.

\subsection{Vlasov dynamics and thermal equilibrium solution}
In the mean-field limit, the phase-space dynamics of the system is governed by two coupled Vlasov equations, one for each species
\begin{equation}\label{Vlasovequations}
    \frac{\partial f_\alpha}{\partial t}+\frac{p}{M_{\alpha}}\frac{\partial f_{\alpha}}{\partial \theta}+F_{\alpha}[f_{\alpha}]\frac{\partial f_{\alpha}}{\partial p}=0, \quad  F_\alpha=-\frac{\partial H_{\alpha}}{\partial \theta} \quad,
\end{equation}
where $f_\alpha$ are the single-particle distribution functions\footnote{From now on, unless explicitly stated, we will refer to $f_{\alpha}$ as the distribution functions} of species $\alpha$, and $F_{\alpha}$ is the mean-field force derived from the Hamiltonian
\begin{equation}\label{Mean-field-Hamiltonians}
        H_\alpha=\frac{p^2}{2M_\alpha}+\mathrm{sign}{(e_\alpha)} C\phi(\theta)+2\tilde{g}\cos{\biggl(\frac{\theta}{2}\biggl)}\,; \quad
        \phi(\theta)= Q \left(\cos{(\theta)}+1 \right)\,,
\end{equation}
where the charge imbalance parameter $Q$ is a functional of the distribution functions
\begin{equation}\label{qVlasov}
        Q[f_{\alpha}]=\sum_{\alpha \in \{e,p\}} \mathrm{sign}{\left(e_{\alpha}\right)} q_{\alpha}[f_{\alpha}], \quad
        q_{\alpha}[f_{\alpha}]=\int_{-\pi}^{\pi} d\theta \int_{-\infty}^{\infty}dp \cos{\left(\theta \right)} f_{\alpha}\left(\theta,p \right) \quad.
\end{equation}
The thermal equilibrium solution corresponds to the isothermal atmosphere, where the distribution function takes the form
\begin{equation}\label{thermalequilibriumsolution}
    f_{\alpha} \left(\theta,p\right)=\frac{e^{-\frac{\tilde{H}_{\alpha}}{T}}}{Z_{\alpha}}, \quad
    Z_{\alpha}=\int_{-\pi}^{\pi} d\theta \int_{-\infty}^{+\infty} dp\,  e^{-\frac{\tilde{H}_{\alpha}}{T}} \quad,
\end{equation}
where $\tilde{H}_{\alpha}$ are the mean-field Hamiltonians \eqref{Mean-field-Hamiltonians} with $\phi=0$, that are
\begin{equation}\label{Stationarymeanfieldhamiltonians}
    \tilde{H}_\alpha=\frac{p^2}{2M_\alpha}+2\tilde{g}\cos{\left(\frac{\theta}{2}\right)} \quad.
\end{equation}

\section{Extended temporal coarse-graining}\label{sec3}

In this section, we develop a general theory of temporal coarse-graining, and subsequently show how it reduces to the specific case presented in \cite{Barbieri2024b}.

\begin{figure}
\centering   
\includegraphics[width=0.99\columnwidth]{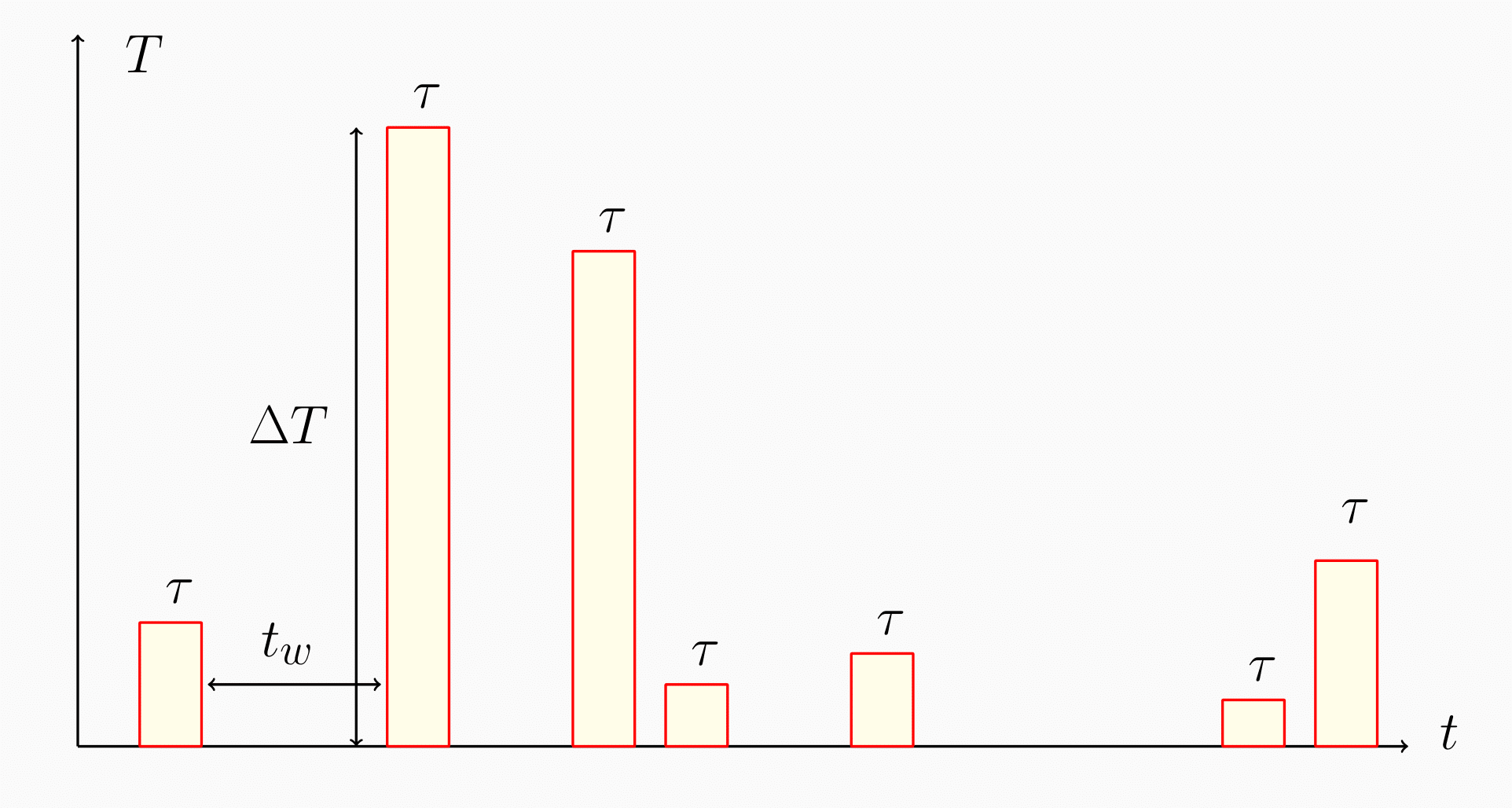}
    \caption{Scheme of the time series of the temperature of the thermal boundary. During the time intervals of duration $\tau$, the temperature increases by an amount $\Delta T$, and during the waiting times $t_w$, it returns to the value $T_0$.
    }
    \label{fig:Temperatureflucscheme}
\end{figure}

\subsection{Extended temporal coarse-graining: kinetic formalism}\label{subsec:temporalcoarsegrainedKT}
If the boundary temperature $T_0$ is held constant, the loop evolves to thermal equilibrium at the same temperature. However, as discussed in (\citealt{barbieri2023temperature}), the chromosphere is highly dynamic. We thus assume its temperature fluctuates due to heating events of amplitude $\Delta T$, drawn from a distribution $\gamma(\Delta T)$, with duration $\tau$, separated by waiting times $t_w$  drawn from a distribution $\eta(t_w)$, during which the boundary returns to temperature $T_0$. A schematic representation of this time-dependent thermal boundary is shown in Fig.\ \ref{fig:Temperatureflucscheme}. In what follows, we present a general formulation of the temporal coarse-graining theory. Specifically, we consider the time average of the Vlasov equations over a coarse-graining interval $\tilde{t}$, such that
\begin{equation}\label{coarsegrainingcondition}
    \tau, \langle t_w \rangle \ll \tilde{t} \quad,
\end{equation}
we also focus on the stationary condition
\begin{equation}
    \biggl \langle \frac{\partial f_{\alpha}}{\partial t} \biggl \rangle_{\tilde{t}} = 0 \,.
\end{equation}
Averaging the Vlasov equation over the time interval $\tilde{t}$ yields
\begin{equation}
    \frac{p}{M_\alpha}\frac{\partial \tilde{f}_{\alpha}}{\partial \theta}+\biggl \langle F_{\alpha}[f_{\alpha}]\frac{\partial f_{\alpha}}{\partial p} \biggl \rangle_{\tilde{t}} = 0 \quad.
\end{equation}
where the coarse-grained distribution function $\tilde{f}_{\alpha}$ is defined via time-averaging as
\begin{equation}\label{coarsegrainedDF}
    \tilde{f}_{\alpha}\left(\theta,p,\tilde{t}\right) = \langle f_{\alpha} \rangle_{\tilde{t}} = \frac{1}{\tilde{t}}\int_{\tilde{t}} dt f_{\alpha}(\theta,p,t)\quad.
\end{equation}
We then decompose the distribution function as $f_\alpha = \tilde{f}_\alpha + \delta f_\alpha$, where $\tilde{f}_\alpha$ is the coarse-grained component and $\delta f_\alpha$ represents the fluctuations around it. Substituting this decomposition into the averaged equation, we obtain
\begin{equation}\label{coarsegraineddynamicspro}
         \frac{p}{M_\alpha}\frac{\partial \tilde{f}_{\alpha}}{\partial \theta}+F_{\alpha}[\tilde{f}_{\alpha}]\frac{\partial \tilde{f}_{\alpha}}{\partial p} = -\mathrm{sign}(e_{\alpha}) C \biggl \langle \delta {E} \frac{\partial \delta f_{\alpha}}{\partial p} \biggl \rangle_{\tilde{t}} \quad.
\end{equation}
Here, $\delta E$ is the fluctuating part of the electrostatic field, given by
\begin{equation}
    \delta E = Q[\delta f_{\alpha}] \sin{(\theta)} \quad Q[\delta f_{\alpha}] = \sum_{\alpha \in \{e,p\}} \mathrm{sign}(e_{\alpha})\int_{-\infty}^{+\infty}dp \int_{-\pi}^{\pi}d\theta \cos{(\theta)} \delta f_{\alpha}(\theta,p).
\end{equation}
Time-averaging the fluctuating boundary conditions leads to a coarse-grained energy reservoir described by
\begin{equation}\label{fluctuatingthermostat}
    \tilde{f}_{\alpha} \left(p\right)=\frac{A}{M_{\alpha}}\int_1^{+\infty}dT\frac{\gamma \left(T\right)}{T}e^{-\frac{p^2}{2TM_{\alpha}}}+\frac{1-A}{M_{\alpha}}e^{-\frac{p^2}{2M_{\alpha}}}\quad,
\end{equation}
with
\begin{equation}\label{fractionoftime}
    A=\frac{\tau}{\tau +\langle t_w \rangle_{\eta}}\quad. 
\end{equation}
We can now split the left-hand side of Eq.\ \eqref{coarsegraineddynamicspro} into two physically distinct contributions:
\begin{itemize}
\item \textbf{Mean-field term (MF)}, describing the standard Vlasov evolution:
\begin{equation}\label{MFterm}
    \biggl(\frac{\partial \tilde{f}_{\alpha}}{\partial \tilde{t}}\biggl)_{MF} = -\frac{p}{M_\alpha}\frac{\partial \tilde{f}_{\alpha}}{\partial \theta}-F_{\alpha}[\tilde{f}_{\alpha}]\frac{\partial \tilde{f}_{\alpha}}{\partial p}\,;
\end{equation}
\item \textbf{Coarse-graining term (CG)}, which originates from the time-averaging of the fluctuations
\begin{equation}\label{collisionalterm}
    \biggl(\frac{\partial \tilde{f}_{\alpha}}{\partial \tilde{t}}\biggl)_{CG}=-\mathrm{sign}(e_{\alpha}) C \biggl \langle \delta {E} \frac{\partial \delta f_{\alpha}}{\partial p} \biggl \rangle_{\tilde{t}}\,.
\end{equation}
\end{itemize}
This second term is absent in the restricted formulation described in \citealt{Barbieri2024b} and introduces corrections that depend explicitly on the electrostatic coupling and on the magnitude of the fluctuations. 

We now aim to estimate under which conditions the coarse-graining correction becomes relevant. To this end, we introduce the characteristic time scale over which the energy of each species is redistributed along the loop, referred to as the relaxation time $t_{R,\alpha}$. This quantity can be approximated as the minimum between the thermal and gravitational crossing times:
\begin{equation}\label{relaxationtimes}
        t_{R,\alpha}\equiv {\rm min}\left(t_{th,\alpha};t_{\tilde{g},\alpha}\right), \qquad
        t_{th}=\sqrt{M_{\alpha}}2\pi, \qquad
        t_{\tilde{g}}=\sqrt{M_{\alpha}}\sqrt{\frac{\pi}{\tilde{g}}}\quad.
\end{equation}

If the heating duration $\tau$ and the waiting time $t_w$ are both much smaller than $t_{R,\alpha}$, we do not expect significant energy fluctuations within the coarse-graining interval $\tilde{t}$, since the system has not sufficient time to redistribute energy along the loop. In this regime, we expect that the time-averaged distribution $\tilde{f}_{\alpha}$ closely approximates the instantaneous distribution $f_{\alpha}$ . Conversely, if $\tau$ and/or $t_w$ are comparable to or larger than, $t_{R,\alpha}$, the system cannot fully equilibrate between heating events, leading to significant energy fluctuations induced by the time-dependent thermal boundary. In this case, $f_{\alpha} \neq \tilde{f}_{\alpha}$ , and the deviation can be estimated as
\begin{equation}\label{estimationfluctuations}
   \delta f_{\alpha} \sim \mathrm{max}\left(\mathcal{O}\left(\frac{\tau}{t_{R,\alpha}}\right),\mathcal{O}\left( \frac{t_w}{t_{R,\alpha}}\right)\right) 
\end{equation}
Given this estimate for the amplitude of fluctuations, we can also assess the magnitude of the coarse-graining correction terms in the kinetic equation
\begin{equation}\label{Ginzburgcriteria}
    \left(\frac{\partial \tilde{f}_{\alpha}}{\partial t}\right)_{CG} \sim \mathrm{max}\left(\mathcal{O}^2\left(\frac{\tau}{t_{R,\alpha}}\right),\mathcal{O}^2\left(\frac{t_w}{t_{R,\alpha}}\right)\right)
\end{equation}
Equation \eqref{Ginzburgcriteria} thus provides a criterion for estimating the relevance of fluctuations, analogous in spirit to the Ginzburg criterion used to assess fluctuation effects near phase transitions.

\subsection{Restricted formulation of the temporal coarse-graining}\label{subsec:restrictedtemporalcoarsegraining}
Using Eq. \eqref{Ginzburgcriteria}, we can estimate the magnitude of the coarse-graining term in the regime
\begin{equation}\label{stationarystateregime}
    \tau, t_w \ll t_{R,e} \quad,
\end{equation}
in which the heating duration and waiting times are both much shorter than the electron relaxation time. In this limit, the coarse-graining correction is negligible, effectively of order zero. As a result, the plasma dynamics can be accurately described by a set of Vlasov equations for the coarse-grained distribution functions
\begin{equation}\label{coarsegraineddynamics}
    \begin{gathered}
         \frac{p}{M_\alpha}\frac{\partial \tilde{f}_{\alpha}}{\partial \theta}+\tilde{F}_{\alpha}[\tilde{f}_{\alpha}]\frac{\partial \tilde{f}_{\alpha}}{\partial p} =0\quad.
    \end{gathered}
\end{equation}
In conclusion, under the condition \eqref{stationarystateregime}, we recover the specific stationary-state regime previously discussed in \cite{Barbieri2024b}. Analytical stationary solutions compatible with the energy reservoir described by Eq. \eqref{fluctuatingthermostat} are
\begin{equation}\label{coarsegrainedstationarysolution}
    \tilde{f}_{\alpha}(\theta,p)=\mathcal{A}_{\alpha}\biggl(A\int_1^{+\infty}dT\frac{\gamma(T)}{T}e^{-\frac{\tilde{H}_{\alpha}}{T}}+(1-A)e^{-\tilde{H}_{\alpha}}\biggl) \quad.
\end{equation}
The constants $\mathcal{A}_{\alpha}$ are normalisation constants, so that $\tilde{f}_{\alpha}$ are normalised to $1$.
The stationary-state distribution functions can be interpreted as the sum of two components: a thermal distribution at the base temperature $T_0=1$ (i.e., the reference temperature set by the thermostat), and a non-thermal contribution arising from the average of thermal distributions at higher temperatures  $T>T_0$, weighted by the probability distribution $\gamma(T)$ of temperature fluctuations. The amplitude of this non-thermal contribution is proportional to $A$, which represents the fraction of time the thermostat deviates from $T_0$ that is, the fraction of time during which the chromosphere is actively heated. At low altitudes $z$, the thermal component dominates; however, its influence decreases with height due to the suppression by the gravitational potential embedded in $H_{\alpha}$. In contrast, the non-thermal component becomes increasingly significant at larger $z$ , where velocity filtration allows higher-energy particles to populate greater heights, manifesting as suprathermal tails in the distribution functions. From these expressions, one can compute the number density and kinetic temperature as moments of $\tilde{f}_{\alpha}$. For instance, the number density is given by
\begin{equation}\label{NEqdensity}
   \tilde{n}_{\alpha}\left(\theta \right)=\frac{A\int_{1}^{+\infty}dT\frac{\gamma \left(T\right)}{\sqrt{T}}e^{-\frac{2\tilde{g}}{T}\cos{\left(\frac{\theta}{2}\right)}}+(1-A)e^{-2\tilde{g}\cos{\left(\frac{\theta}{2}\right)}}}{A\int_{1}^{+\infty}dT\frac{\gamma \left(T\right)}{\sqrt{T}}\int_{-\pi}^{\pi} d\theta e^{-\frac{2\tilde{g}}{T}\cos{\left(\frac{\theta}{2}\right)}}+(1-A)\int_{-\pi}^{\pi} d\theta e^{-2\tilde{g}\cos{\left(\frac{\theta}{2}\right)}}}\quad,
\end{equation}
and the kinetic temperature is
\begin{equation}\label{NEqtemperature}
    \tilde{T}_{\alpha}(\theta)=\frac{A\int_{1}^{+\infty}dT\gamma \left(T\right)\sqrt{T}e^{-\frac{2\tilde{g}}{T}\cos{\left(\frac{\theta}{2}\right)}}+(1-A)e^{-2\tilde{g}\cos{\left(\frac{\theta}{2}\right)}}}{A\int_{1}^{+\infty}dT\frac{\gamma \left(T\right)}{\sqrt{T}}e^{-\frac{2\tilde{g}}{T}\cos{\left(\frac{\theta}{2}\right)}}+(1-A)e^{-2\tilde{g}\cos{\left(\frac{\theta}{2}\right)}}}\quad.
\end{equation}
It is important to note that in this regime, both the density and temperature profiles are the same for electrons and protons, and are independent of the electrostatic coupling parameter $C$. However, this symmetry breaks in the extended formulation as discussed above. This implies that, when $C \neq 0$, the stationary density and temperature profiles of electrons and protons may differ, and the mechanism of gravitational filtering may no longer guarantee temperature inversion.

In the next section, we will develop a two-fluid formalism based on this extended kinetic description in order to assess the role of electrostatic fluctuations more precisely.

\subsection{Extended temporal coarse-graining: two-fluid formalism}\label{coarsegrainedtwofluidformalism}
We can derive a set of two-fluid equations directly from the coarse-grained kinetic equations established in Section \ref{subsec:temporalcoarsegrainedKT}. The method relies on computing the moments of the distribution functions. Specifically, for any function of momentum $\psi(p)$, we define the moment
\begin{equation}\label{moments}
    \langle \psi(p) \rangle_p = \int_{-\infty}^{+\infty}dp \, \psi(p)\, \tilde{f}_{\alpha}(\theta,p,\tilde{t}) \quad .
\end{equation}
Multiplying the kinetic equation by $\psi(p)$ and integrating over $p$, we obtain a hierarchy of fluid-like equations
\begin{equation}
    \frac{\partial}{\partial \theta} \biggl \langle \psi(p) \frac{p}{M_{\alpha}} \biggl \rangle_p - \biggl \langle  F_{\alpha} \frac{\partial \psi(p)}{\partial p}\biggl \rangle_p = \biggl \langle \psi (p) \biggl(\frac{\partial f_{\alpha}}{\partial t}\biggl)_{CG} \biggl \rangle_p \quad.
\end{equation}
For $\psi(p) = 1$, we obtain the continuity equation
\begin{equation}\label{coarsegrainedcontinuity}
    \frac{\partial (\tilde{n}_{\alpha} \tilde{u}_{\alpha})}{\partial \theta} =0 \quad,
\end{equation}
where the coarse-grained density $n_\alpha$ and mean velocity $u_\alpha$ are defined as
\begin{equation}\label{coarsegraineddensityandvelocityfield}
    \tilde{n}_{\alpha}(\theta,\tilde{t})=\langle1\rangle_p=\int_{-\infty}^{+\infty}dp \tilde{f}_{\alpha}(\theta,p,\tilde{t}) \qquad \tilde{u}_{\alpha}(\theta,\tilde{t})=\frac{1}{M_{\alpha}}\frac{\langle p \rangle_p}{\tilde{n}_{\alpha}} \quad.
\end{equation}
For $\psi(p) = p / M_\alpha$, we obtain the momentum balance equation
\begin{equation}\label{two-fluidEqmotion}
    \frac{\partial}{\partial \theta}\biggl(M_{\alpha}\tilde{n}_{\alpha} \tilde{u}_{\alpha}^2 + \tilde{P}_{\alpha}\biggl)=\tilde{n}_{\alpha}F_{\alpha}[\tilde{f}_{\alpha}]+ R_{\alpha}  \quad,
\end{equation}
where $P_\alpha$ is the coarse-grained kinetic pressure of species $\alpha$, defined as
\begin{equation}\label{kineticpressure}
    \tilde{P}_{\alpha} = M_{\alpha}\biggl \langle \biggl(\frac{p}{M_{\alpha}}-\tilde{u}_{\alpha}\biggl)^2\biggl\rangle_p \quad,
\end{equation}
and $R_\alpha$ is the contribution from the coarse-graining correction term
\begin{equation}
    R_{\alpha} = \biggl \langle p \biggl( \frac{\partial \tilde{f}_{\alpha}}{\partial \tilde{t}}\biggl)_{CG} \biggl \rangle_p = -\mathrm{sign}(e_{\alpha}) C \biggl\langle \delta E \int_{-\infty}^{+\infty}dp \delta f_{\alpha} \biggl \rangle_{\tilde{t}} \quad.
\end{equation}
This additional term $R_\alpha$ quantifies the strength of the fluctuations due to the electrostatic field, and provides a way to measure the influence of coarse-graining on the macroscopic momentum balance. In regimes where $R_\alpha$ is non-zero, the coarse-graining term modifies the dynamics, leading to deviations from the standard Vlasov-based mean-field predictions. In Section \ref{subsec:hydridstateregime}, we will use this two-fluid formalism in conjunction with numerical simulations to evaluate $R_\alpha$ and examine how the presence of fluctuations changes the stationary profiles of density and temperature across different regimes.

\subsection{Superposition of many thermal solutions}\label{subsec:Superpositiontheory}
We now consider the opposite regime with respect to the restricted coarse-graining scenario presented in Section \ref{subsec:restrictedtemporalcoarsegraining}. Here, we assume that the system has sufficient time to relax to thermal equilibrium during each heating or cooling phase. That is, the characteristic relaxation times $t_{R,\alpha}$ are much shorter than both the duration of heating events $\tau$ and the average waiting time between them $\langle t_w \rangle$,  that is
\begin{equation}
 \tau,\langle t_w \rangle \gg t_{R,p}\,.  
\end{equation}
This corresponds to a regime where the plasma continuously oscillates between well-defined thermal equilibria.
In this regime, the long-time dynamics of the system, on timescales much larger than $\tau$ and $\langle t_w \rangle$, can be described by a coarse-grained distribution that results from averaging over the sequence of these thermal states. The effective distribution is then given by
\begin{equation}\label{VDFsuperpositiontemperatures}
    \tilde{f}_{\alpha,SS}(\theta,p)=A\int_{1}^{+\infty}dT\, \gamma(T)\, f_{\alpha,T}(\theta,p) +(1-A)f_{\alpha,1}(\theta,p) \,,
\end{equation}
where $f_{\alpha,T}(\theta,p)$ is the thermal solution at the temperature $T$ computed via Eq.\ \eqref{thermalequilibriumsolution}. These distribution functions exhibit suprathermal tails, arising from the superposition of multiple thermal equilibrium states\footnote{A distribution function formed as a superposition of thermal states is commonly referred to as a superstatistics; see \citealt{BECKsuper}.} contained in the first term of Eq.~\eqref{VDFsuperpositiontemperatures}. 
As a result, the density and temperature profiles derived from such distributions necessarily exhibit temperature inversion, a signature feature of gravitational filtering of higher-energy tails. The inverted profiles can be explicitly calculated analytically and we get for the number density
\begin{equation}\label{superpositiondensities}
\tilde{n}_{SS}(\theta)=A\int_1^{+\infty}dT \, \gamma(T) n_{T}(\theta)+(1-A)\, n_{1}(\theta)\,,
\end{equation}
where $n_{T}(\theta)$ is given by
\begin{equation}
    n_{T}(\theta)=\frac{e^{-\frac{2 \tilde{g}}{T}\cos{\bigl(\frac{\theta}{2 }\bigl)}}}{\int_{-\pi}^{\pi}d\theta\, e^{-\frac{2 \tilde{g}}{T}\cos{\bigl(\frac{\theta}{2}\bigl)}}}\,.
\end{equation}
while for the kinetic temperature we get
\begin{equation}\label{superpositiontemperatures}
    \tilde{T}_{SS}(\theta) = \frac{A\int_1^{+\infty}dT\, \gamma(T)\, T^{3/2} n_{T}(\theta)+(1-A) n_{1}(\theta)}{A\int_1^{+\infty}dT\, \gamma(T)\, n_{T}(\theta)+(1-A) n_{1}(\theta)}\,.
\end{equation}
It is important to emphasize that the temperature obtained from Eq.~\eqref{superpositiontemperatures} is the kinetic temperature of the time-averaged distribution function: since the system evolves through a sequence of distinct thermal states, this expression does not represent the time-average of the individual equilibrium temperatures. The latter, which characterizes the mean energy content of the sequence of thermal states, is given by:
\begin{equation}\label{superpositiontemperaturestrue}
    T_{SP} = A \int_{1}^{+\infty}dT\, \gamma(T)\, T +(1-A) \,.
\end{equation}
We note that, just as in the opposite regime described in Section~\ref{subsec:restrictedtemporalcoarsegraining}, the coarse-grained temperatures computed via Eqs.\ \eqref{superpositiontemperatures} and \eqref{superpositiontemperaturestrue}, as well as the number densities obtained from Eq.\ \eqref{superpositiondensities}, are independent of the self-consistent electrostatic interaction, and are thus identical for both species.

\section{Dynamics of the plasma model}\label{sec4}

We now apply the coarse-graining formalism to the model of a gravitationally confined plasma described in Section \ref{sec2}.
\cite{Barbieri2024b} demonstrated that the plasma develops a hot corona, i.e., relaxes towards a non-thermal stationary configuration exhibiting temperature inversion, if two key conditions are met:
\begin{itemize}
    \item The timescales of thermal fluctuations ($\tau$ and $t_w$) must be much shorter than the electron relaxation time;
    \item The heating events must be rare and intense, i.e., $\tau \ll \langle t_w \rangle$ and $\Delta T \gg T_0$.
\end{itemize}
In what follows, we explore other dynamical regimes, i.e., we relax the condition $\tau, t_w \ll t_{R,e}$, while keeping the constraints on $\Delta T$ and $\langle t_w \rangle$. For clarity, we present a simplified case in which the boundary temperature alternates between two fixed values, $T = 1$ and $T = 1 + \Delta T$, with fixed waiting time $t_w$. The associated distributions are
\begin{equation}
    \gamma(T)=\delta(T-(1+\Delta T))\,; \quad \eta(t_w)=\delta(t-t_w) \, .
\end{equation}
Although this case is specific, the physical conclusions that follow are general.

\subsection{Computation of the relevant quantities}\label{subsec:computationmoments}

To characterize the dynamical evolution of the plasma system, we compute both the kinetic energy and the stratification parameter for each species. These quantities are given by
\begin{equation}\label{Kineticandstrat}
    K_{\alpha}=\frac{1}{N}\sum_{j=1}^N \frac{p_{j,\alpha}^2}{2M_{\alpha}}, \quad q_{\alpha}=\frac{1}{N}\sum_{j=1}^N \cos{(\theta_{j,\alpha})} \quad.
\end{equation}
To compute the moments of the distribution functions of a physical quantity $\psi(p)$ given by \eqref{coarsegrainedDF}, we use a two-step time-coarse-graining procedure: Furthermore, the moments of the distribution functions of a physical quantity $\psi(p)$ have been computed at a temporal coarse-graining level as follows
\begin{enumerate}
\item Instantaneous moments of the distribution functions are computed at each time step.
\item These moments are then time-averaged over a coarse-graining window $\tilde{t} \gg \tau,t_w$, ensuring smoothing over thermal boundary fluctuations.
\end{enumerate}
Using the definition of the coarse-grained distribution functions in Eq.\ \eqref{coarsegrainedDF}, this procedure recovers the kinetic moments defined in Eq.\ \eqref{moments}, ensuring that all coarse-grained quantities depend solely on the time-averaged distributions. 

As for temperature, since it is defined through the ratio of two moments, there is no single way to define it. In particular, we can define temperature at the level of temporal coarse-graining in two different ways. 
\begin{enumerate}
    \item The first defines temperature by time-averaging the instantaneous kinetic temperature
    \begin{equation}\label{coarsegrainedtempsuper}
    T_{ST,\alpha}(\theta,\tilde{t})=\frac{1}{\tilde{t}}\int_{\tilde{t}} T_{\alpha}(\theta,t) dt \qquad T_{\alpha}(\theta,t)=\frac{P_{\alpha}(\theta,t)}{n_{\alpha}(\theta,t)} \quad,
    \end{equation}
    where $P_{\alpha}$ is kinetic pressure and $n_{\alpha}$ is the number density at time $t$. Here we  assume that the measurements of the moments can be done with a resolution much shorter than the coarse-graining time-scale allowing to recover the microscopic dynamic of the system.
    \item If the dynamic of the system below the coarse-graining time-scale is not accessible by the measurements it is still possible to define an averaged temperature as the ratio of the time-averaged pressure and density
    \begin{equation}\label{coarsegrainedtemp}
    \tilde{T}_{\alpha}(\theta,\tilde{t}) = \frac{\tilde{P}_{\alpha}(\theta,\tilde{t})}{\tilde{n}_{\alpha}(\theta,\tilde{t})} \quad.
    \end{equation}   
    From a physical point of view, this definition is meaningful when the system dynamics below the coarse-graining time scale $\tilde{t}$ is not accessible. In this regime, the behaviour of the two species is described by the coarse-grained distribution functions $\tilde{f}_{\alpha}$, and the kinetic temperature is defined using the kinetic formulation applied to $\tilde{f}_{\alpha}$, as specified by the present definition.
\end{enumerate}
Both definitions are meaningful in the coarse-graining context, but they may yield different results depending on the magnitude of fluctuations $\delta f_{\alpha}$. Indeed, by decomposing the distribution function as $f_{\alpha} = \tilde{f}_{\alpha} + \delta f_{\alpha}$, it is possible to establish a link between the two definitions of temperature given by Eqs.\ \eqref{coarsegrainedtempsuper} and \eqref{coarsegrainedtemp}. After some algebraic manipulation, we obtain
\begin{equation}\label{temperatureimbalance}
    T_{ST,\alpha}=\tilde{T}_{\alpha}\cdot \mathcal{K}_{\alpha}[\delta f_{\alpha}] \qquad \mathcal{K}_{\alpha}[\delta f_{\alpha}]=\biggl(1+\sum_{k=1}^{+\infty}(-1)^k\biggl(\frac{\delta n_{\alpha}}{\tilde{n}_{\alpha}}\biggl)^k\biggl)\cdot\biggl(1+\frac{\delta P_{\alpha}}{\tilde{P}_{\alpha}}\biggl) \quad,
\end{equation}
where $\delta n_{\alpha}$ is given by
\begin{equation}
    \delta n_{\alpha}=\int_{-\infty}^{+\infty}dp \delta f_{\alpha} \quad,
\end{equation}
$\delta P_{\alpha}$ is given by
\begin{equation}
    \delta P_{\alpha} = \frac{1}{M_{\alpha}}\int_{-\infty}^{+\infty}dp p^2 \delta f_{\alpha}-M_{\alpha}(\tilde{n}_{\alpha}\delta u_{\alpha}^2+\tilde{n}_{\alpha}\tilde{u}_{\alpha}\delta u_{\alpha}+\tilde{u}_{\alpha}^2\delta n_{\alpha}+\delta n_{\alpha} \delta \tilde{u}_{\alpha}^2+2\tilde{u}_{\alpha}\delta n_{\alpha}\delta u_{\alpha}) \,, 
\end{equation}
and $\delta u_{\alpha}$ is given by
\begin{equation}
    \delta u_{\alpha}=\frac{\int_{-\infty}^{+\infty}dpp\delta f_{\alpha}}{M_{\alpha}\tilde{n}_{\alpha}}\sum_{k=0}^{+\infty}(-1)^k \biggl(\frac{\delta n_{\alpha}}{\tilde{n}_{\alpha}}\biggl)^k \,.
\end{equation}
The functional $\mathcal{K}_{\alpha}[\delta f_{\alpha}]$ measures the ratio between the two temperature definitions. Furthermore, we have that $\mathcal{K}_{\alpha}[\delta f_{\alpha}]\rightarrow 1$ when $\delta f_{\alpha} \rightarrow 0$. This implies that the two temperature definitions are equal only if the fluctuations are negligible. This difference becomes important in regimes where fluctuations are non-negligible and will be discussed in detail in the subsequent sections.

\subsection{The numerical simulation}\label{subsec:numericalsimulations}

The $N$-particle dynamics described by Eq. \eqref{HMF2Sdimensionless} are solved numerically using a fourth-order symplectic integrator (\citealt{Candy1991}). The thermal flux from the fluctuating boundary is modeled using a standard approach (see \citealt{ThermalWalls,Landi-Pantellini2001}). Specifically, when a particle interacts with the thermal boundary, it is re-injected with a momentum 
$p$ sampled from a flux-weighted distribution:
\begin{equation}
    r = \frac{1}{M_{\alpha}T} \int_0^{p} dx\, x\, e^{-\frac{x^2}{2M_{\alpha}T}}\quad,  
\end{equation}
where $r \in [0,1]$ is a uniformly distributed random variable, $T$ denotes the instantaneous temperature of the thermal boundary and $p$ denotes the momentum of the particle. We note that naively reintroducing the particle with a new velocity drawn from a half-Gaussian distribution at temperature $T$ would violate the stationary thermal equilibrium, as re-injected particles would, by construction, have a higher probability of possessing velocities close to zero.

To impose central symmetry, we restrict the simulation domain to $\theta \in [-\pi,0]$. When a particle reaches the upper boundary at $\theta = 0$, central symmetry is enforced by applying the transformation
\begin{equation}\label{centralsymmetry}
        \mathrm{if}  \quad \theta_{j,\alpha}\left(t\right)>0: \qquad
        \theta_{j,\alpha}\left(t\right) \rightarrow -\theta_{j,\alpha}\left(t\right), \qquad
        p_{j,\alpha} \rightarrow -p_{j,\alpha}\quad.
\end{equation}

\begin{figure}
    \centering
    \includegraphics[width=0.99\columnwidth]{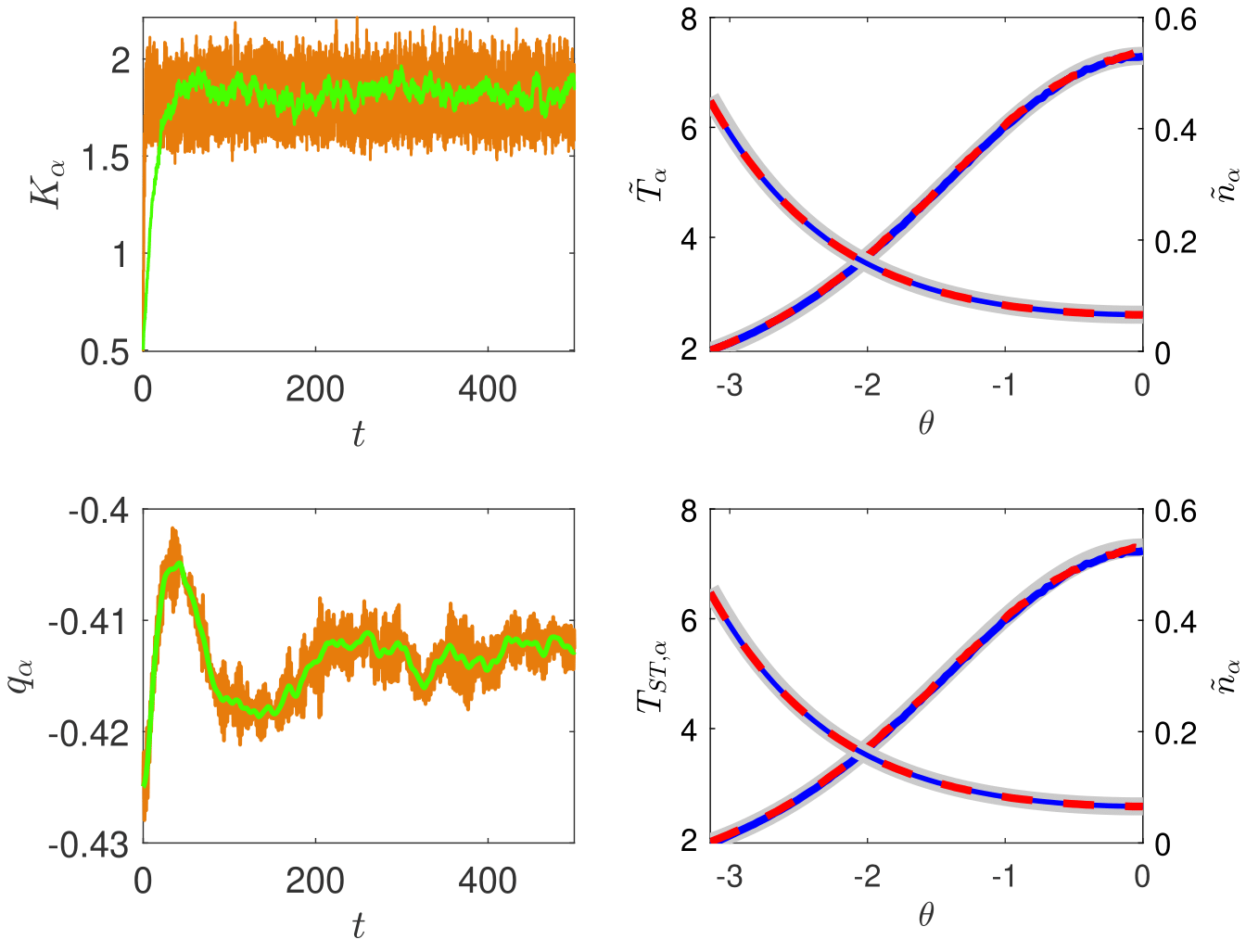}
    \caption{Left panels: time evolution of the kinetic energies $K_{\alpha}$ (top) and stratification parameters $q_{\alpha}$ (bottom) for protons (green) and electrons (orange), computed from simulations using Eq.\ \eqref{Kineticandstrat}. Right panels: Time-averaged temperature and density profiles for electrons (red dashed lines) and protons (blue lines). The upper panel shows the coarse-grained temperatures calculated using Eq.\ \eqref{coarsegrainedtemp}, while the lower panel uses the kinetic temperature definition from Eq.\ \eqref{coarsegrainedtempsuper}. Theoretical predictions from Eq. \eqref{NEqdensity} (for density) and Eq. \eqref{NEqtemperature} (for temperature) are shown as grey lines.
    }
    \label{TempInvQSS}
\end{figure}

\subsection{Stationary-state regime}\label{subsec:stationarystateregime}
As previously discussed, when the timescales associated with temperature fluctuations, namely, $\tau$  and $t_w$ , satisfy the condition
\begin{equation}\label{stationarystatecondition}
    \tau,\langle t_w \rangle \ll t_{R,e} \quad,
\end{equation}
the plasma undergoes a relaxation process towards a non-thermal stationary configuration characterized by inverted density and temperature profiles.
Results from a simulation in this regime are shown in Fig.\ \ref{TempInvQSS}. 
To generate these results, simulation parameters were fixed as follows: $\tau=0.05, t_w = 0.5, g=1, M=100, C=400, \Delta T=90, N=2^{16}$, and time step $dt=4\cdot 10^{-3}$. The left panels in Fig.\ \ref{TempInvQSS} illustrate the temporal evolution of the kinetic energies $K_{\alpha}$ and the stratification parameters $q_{\alpha}$ for both electrons and protons. As shown, both quantities tend towards a stationary regime. The coarse-grained temperature and density profiles, obtained by time averaging during the steady phase, as described in Section \ref{subsec:computationmoments}, are shown in the right panels of Fig.\ \ref{TempInvQSS}. Theoretical predictions were computed using Eq.\ \eqref{NEqtemperature} for the temperature profiles and Eq.\eqref{NEqdensity} for the number densities.
The top right panel shows temperatures computed according to the definition in Eq.\ \eqref{coarsegrainedtemp}, while the bottom right panel uses the definition in Eq.\ \eqref{coarsegrainedtempsuper}. In both cases, the simulations produce identical results for both species, in excellent agreement with the theoretical predictions. This agreement stems from the fact that fluctuations are negligible in this regime. As a result, according to Eq.\ \eqref{temperatureimbalance}, the two definitions of coarse-grained temperature must coincide. In this regime, the general temporal coarse-graining formalism described in Section \eqref{subsec:temporalcoarsegrainedKT} simplifies substantially. Specifically, the fluctuation term $\delta f_{\alpha}$ is negligible and the additional coarse graining term in the kinetic equation \eqref{collisionalterm} vanishes. Consequently, the system reduces to the stationary form of the Vlasov equation, and the stationary solutions are those given by Eq.\ \eqref{coarsegrainedstationarysolution}. Remarkably, these solutions are independent of the self-consistent electrostatic interaction parameter $C$, as confirmed by the numerical simulations.

\subsection{Hybrid state regime}\label{subsec:hydridstateregime}
By increasing the time scales $\tau$ and $t_w$ the system reaches the following regime:
\begin{equation}\label{hybridstatecondition}
    \tau < t_{R,e} < \langle t_w \rangle < t_{R,i} \,.
\end{equation}
In this regime, electrons do not have sufficient time to reach thermal equilibrium at temperature $T = 1 + \Delta T$ during a heating event of duration $\tau$, but they do relax to the base temperature $T_0 = 1$ during the longer waiting time $t_w$. Protons, by contrast, remain in the regime described in the previous section, since both $\tau$ and $t_w$ are still much smaller than their relaxation time $t_{R,i}$. We refer to this scenario as the ``hybrid state regime''.
\begin{figure}
    \centering
    \includegraphics[width=0.99\columnwidth]{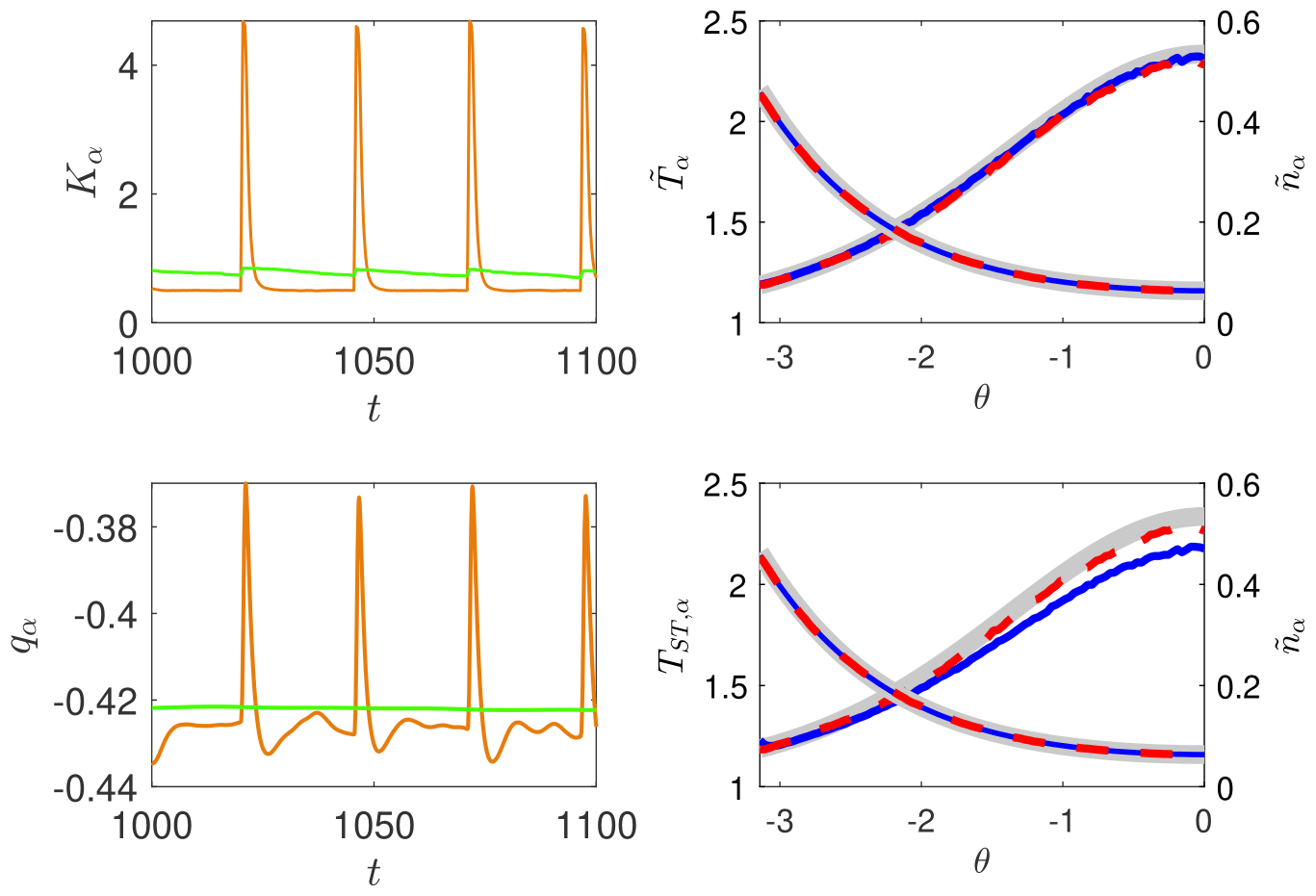}
    \caption{Same quantities and color scheme as in Fig.\ \ref{TempInvQSS}, but in the hybrid regime described in Sec.\ \ref{subsec:hydridstateregime}, where electron and proton dynamics exhibit different relaxation behaviours.}   
    \label{TempInvHS}
\end{figure}
Numerical simulations in this regime reveal that the plasma dynamics depends sensitively on the presence or absence of self-consistent electrostatic interactions. Accordingly, we first analyze the case $C = 0$ before introducing the effects of a finite $C$.
\subsubsection{Independent atmospheres}
We begin by examining the dynamics in the absence of self-consistent electrostatic interactions, i.e., with $C = 0$. 
For the simulations shown in Figure \ref{TempInvHS}, the model parameters were set as follows: $\tau = 0.5$, $t_w = 25$, $\tilde{g} = 1$, $M = 1836$, $C = 0$, $\Delta T = 90$, $N = 2^{16}$, and time step $dt = 4 \cdot 10^{-3}$. The left panels of Figure \ref{TempInvHS} show that both the electron kinetic energy and the stratification parameter exhibit picks associated with the heating events of duration $\tau$. During the waiting time $t_w \gg t_{R,e}$, the electrons relax back toward a thermal configuration at temperature $T_0 = 1$. As a result, the electron dynamics is no longer truly stationary, in contrast to the protons, which remain effectively stationary due to $\tau, t_w \ll t_{R,i}$. Nevertheless, when observed on a coarse-grained timescale $\tilde{t}$ as defined in Eq.\eqref{coarsegrainedtemp} are identical for electrons and protons, and also match the analytical prediction given by Eq. \eqref{NEqtemperature}.
This can be understood as follows. First of all $C=0$ and in addition the coarse-grained dynamics is stationary so that Eq. \eqref{stationarystatecondition} is satisfied. These conditions imply that the general kinetic equation given in Eq.\eqref{coarsegraineddynamicspro} reduces to a stationary system of Vlasov equations for the two species, as in the steady-state regime described in Sec. \ref{subsec:stationarystateregime}. Consequently, the temperature profiles computed via Eq.\eqref{coarsegrainedtemp} must be identical for both species and match the theoretical profile given in Eq.\eqref{NEqtemperature}. Moreover, this also justifies the equality of the coarse-grained number densities for electrons and protons, which coincide with the analytical profile from Eq.\eqref{NEqdensity}. These results are fully consistent with the temporal coarse-graining two-fluid formalism presented in Section\ref{coarsegrainedtwofluidformalism}. In this specific case, the two-fluid equations of motion, Eq.~\eqref{two-fluidEqmotion}, reduce to the hydrostatic balance:
\begin{equation}\label{coarsegrainedhydrostatic}
    \frac{\partial \tilde{P}_{\alpha}}{\partial \theta} = -\tilde{n}_{\alpha}\tilde{g}\sin{\biggl(\frac{\theta}{2}\biggl)} \quad.
\end{equation}
Given the condition $\tilde{n}_e = \tilde{n}_p$, we can subtract the equations for the two species to obtain
\begin{equation}
    \sum_{\alpha \in \{e,p\}} \mathrm{sign}(e_{\alpha}) \frac{\partial \tilde{P}_{\alpha}}{\partial \theta }  = 0 \quad.
\end{equation}
This relation admits the equality of coarse-grained pressures as a valid solution. Combining the equalities of pressure and density, we conclude that the coarse-grained temperatures must also be equal, i.e., $\tilde{T}_e = \tilde{T}_p$. However, since the electron dynamics is no longer locked in the non-equilibrium stationary state, the condition $\tilde{f}_e = f_e$ no longer holds, and the fluctuation component $\delta f_e$ becomes non-zero. According to Eq.~\eqref{temperatureimbalance}, this implies a mismatch between the two definitions of temperature, so that $T_{ST,e} \neq \tilde{T}_e$. This difference is clearly visible in the right panels of Fig.~\ref{TempInvHS}.

\subsubsection{The role of the self-electrostatic interaction}
\begin{figure}
    \centering
    \includegraphics[width=0.99\columnwidth]{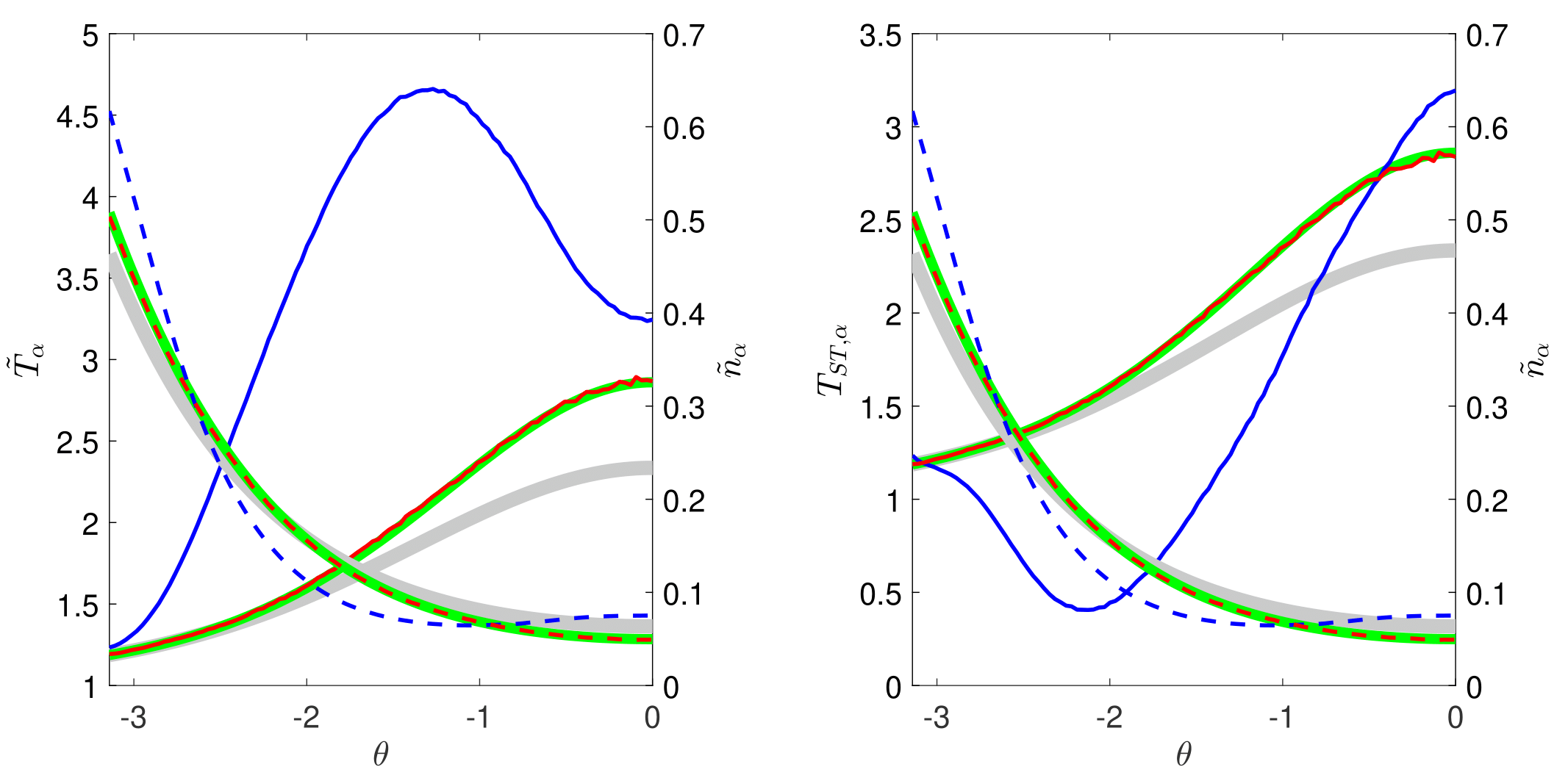}
    \caption{Left panel: Coarse-grained density (dotted lines) and temperature (solid lines) profiles for protons (red) and electrons (blue), calculated using Eq. \eqref{coarsegrainedtemp}.Right panel: Same quantities, but temperature profiles are computed using Eq. \eqref{coarsegrainedtempsuper}.In both panels, grey curves show theoretical predictions from Eq. \eqref{NEqdensity} (density) and Eq. \eqref{NEqtemperature} (temperature). Green curves represent proton-only theoretical profiles from Eq. \eqref{NEqdensitywithC} (density) and Eq. \eqref{NEqtemperaturewithC} (temperature), which account for the electrostatic contribution.
    }
    \label{TempInvwithC2}
\end{figure}

Figure~\ref{TempInvwithC2} displays the temperature profiles (solid lines) and number density profiles (dashed lines) obtained from numerical simulations. The blue curves correspond to electrons, while the red curves represent protons. In the left panel, the temperatures are calculated using Eq.\eqref{coarsegrainedtemp}, while in the right panel they are calculated using Eq.\eqref{coarsegrainedtempsuper}. The theoretical temperature and density profiles are plotted in thick gray lines: the decreasing density profile is given by Eq.\eqref{NEqdensity}, and the increasing temperature profile by Eq.\eqref{NEqtemperature}. The simulation parameters were fixed as follows: $\tau = 0.5$, $t_w = 25$, $\tilde{g} = 1$, $M = 1836$, $C = 400$, $\Delta T = 90$, $N = 2^{16}$, and the integration time step was set to $dt = 4 \cdot 10^{-3}$. As evident from Figure~\ref{TempInvwithC2}, the coarse-grained temperature and density profiles no longer agree with the analytical predictions of Eqs.\eqref{NEqdensity} and \eqref{NEqtemperature}, for both species. Moreover, the electron profiles, computed using both Eq.\eqref{coarsegrainedtemp} and Eq.~\eqref{coarsegrainedtempsuper}, no longer exhibit a monotonic increase in temperature with respect to the spatial coordinate $\theta$. These results suggest that, in this regime, the electrostatic interaction plays a non-negligible role in shaping the temperature and density profiles of the plasma. The breakdown of the agreement with the analytical profiles is a direct consequence of the presence of non-negligible fluctuations, whose effects will be analyzed in detail later in this section.

Since $C \neq 0$, the ``coarse-graining'' correction terms described by Eqs.\eqref{collisionalterm} are non-zero. As a result, the system of kinetic equations can no longer be described solely by the mean-field Vlasov dynamics, represented by Eqs.\eqref{MFterm}. Consequently, the distribution functions given by Eq. \eqref{coarsegrainedstationarysolution} are no longer valid solutions of the full kinetic equations. To assess the presence and magnitude of fluctuations, we make use of the two-fluid formalism introduced in Subsection \ref{coarsegrainedtwofluidformalism}. In particular, the momentum-balance equations \eqref{two-fluidEqmotion} are central to this analysis. Due to the assumption of central symmetry, all odd moments of the coarse-grained distribution functions vanish. This property has been confirmed via numerical simulations. As a result, only the pressure gradient terms contribute to the left-hand sides of Eqs.~\eqref{two-fluidEqmotion}. To quantify the intensity of the fluctuations $\delta f_{\alpha}$, we compute the residual terms $R_{\alpha}$ appearing in Eqs.~\eqref{two-fluidEqmotion}. This can be done by evaluating, from simulations, the pressure term as:
\begin{equation}\label{F_1}
    \mathcal{F}_{\alpha,1}= \frac{\partial \tilde{P}_{\alpha}}{\partial \theta} \quad,
\end{equation}
the total force term as
\begin{equation}\label{F_2}
\mathcal{F}_{\alpha,2}=\tilde{n}_{\alpha}F_{\alpha}[\tilde{f}_{\alpha}] \quad.
\end{equation}
Therefore, the difference of Eq. \eqref{F_1} and Eq. \eqref{F_2} gives the ``coarse-graining" force
\begin{equation}\label{F_3}
\mathcal{F}_{\alpha,3}=R_{\alpha} \quad.
\end{equation}

\begin{figure}
    \centering
    \includegraphics[width=0.99\columnwidth]{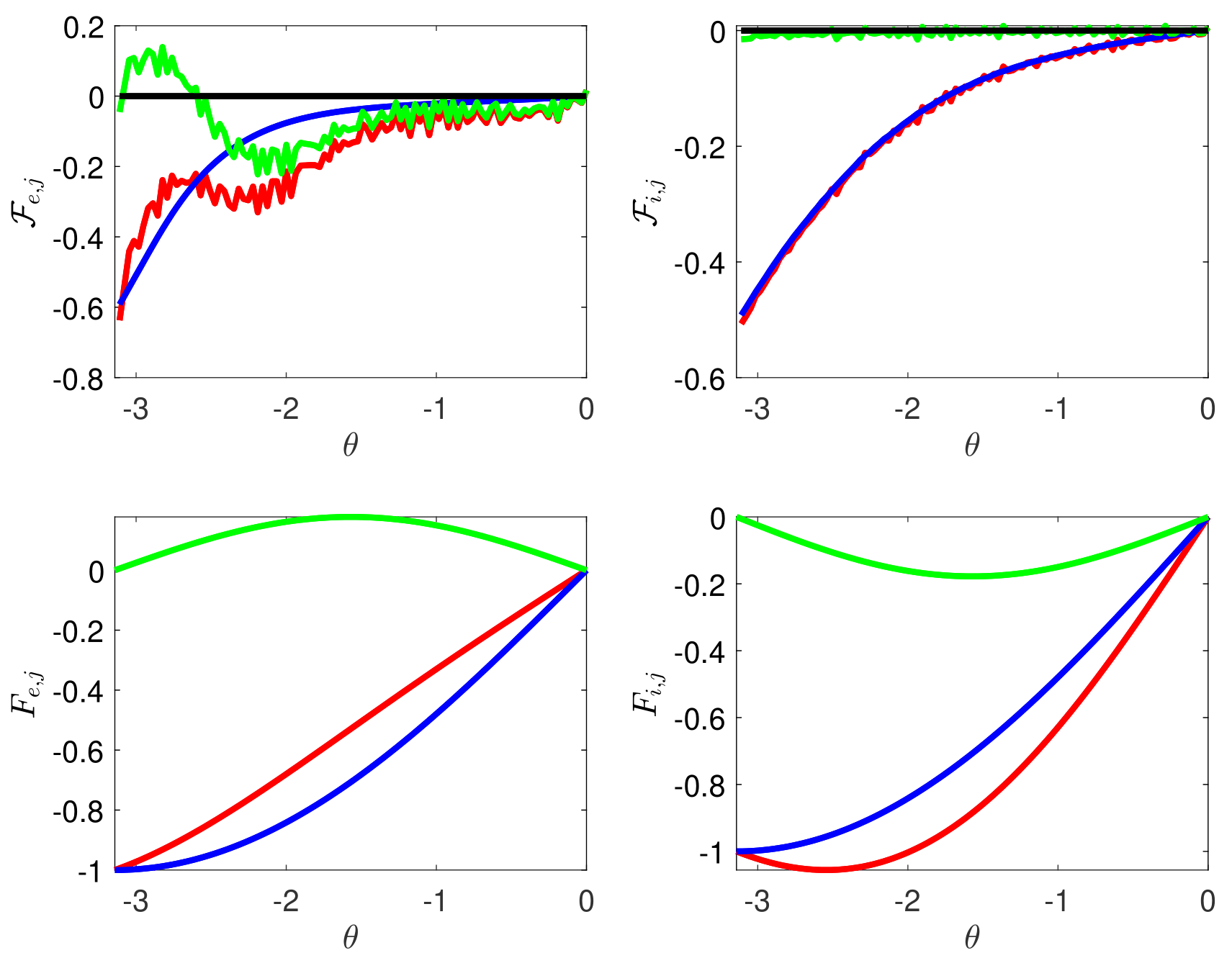}
    \caption{Top left: Electron pressure gradient $\mathcal{F}_{e,1}$ (red), total force  $\mathcal{F}_{e,2}$ (blue), and coarse-graining correction $\mathcal{F}_{e,3}$ (green), as functions of the spatial coordinate $\theta$. The black curve shows the total force balance, which vanishes as expected. Top right: Same quantities for protons. Bottom left: Electrostatic force (green), gravitational force (blue), and total force (red) acting on electrons, calculated respectively from Eqs. \eqref{Elforce}, \eqref{Extforce}, and \eqref{Totforce}. Bottom right: Same as left, but for protons.
    }
    \label{Compareforcesmix}
\end{figure}

In Figure~\ref{Compareforcesmix}, we plot the three force components $\mathcal{F}_{\alpha,j}$ for $j = 1,2, 3$ as functions of $\theta$: the top left panel corresponds to electrons, and the top right to protons. For electrons, it is evident that the pressure gradient force $\mathcal{F}_{\alpha,1}$ (shown in red) differs from the mean-field force $\mathcal{F}_{\alpha,2}$ (blue curve). As a result, the coarse-graining correction $\mathcal{F}_{\alpha,3}$ (green) is non-zero and of comparable magnitude to the other contributions, indicating the presence of significant fluctuations in this regime. In contrast, for protons, since $\tau, \langle t_w \rangle \ll t_{R,p}$, the system remains effectively in the stationary-state regime. Accordingly, we observe that $\mathcal{F}_{\alpha,1} = \mathcal{F}_{\alpha,2}$, implying that the coarse-graining term $\mathcal{F}_{\alpha,3}$ vanishes. The sum of all three force components, shown in black in both panels, satisfies the force balance condition imposed by Eqs.~\eqref{two-fluidEqmotion}, and thus must be identically zero.

Now that the intensity of the fluctuations has been quantified, we shall return to the kinetic approach. Using the Eqs.  \eqref{coarsegraineddynamicspro} and imposing stationarity, we obtain the following kinetic equation
\begin{equation}\label{kineticequationelectrons}
       p\frac{\partial \tilde{f}_{e}}{\partial \theta}+F_{e}[\tilde{f}_{e}]\frac{\partial \tilde{f}_{e}}{\partial p} = C \biggl \langle \delta {E} \frac{\partial \delta f_{e}}{\partial p} \biggl \rangle_{\tilde{t}},
\end{equation}
for the electrons and 
\begin{equation}\label{kineticequationprotons}
       \frac{p}{M}\frac{\partial \tilde{f}_{i}}{\partial \theta}+F_{i}[\tilde{f}_{i}]\frac{\partial \tilde{f}_{i}}{\partial p} = 0,
\end{equation}
for the protons. In the equations discussed above, the coarse-graining correction term due to fluctuations has been retained for electrons but omitted for protons. As a result, the electron temperature and density profiles cannot coincide with the analytical mean-field predictions given by Eq.\eqref{NEqtemperature} for temperature and Eq.\eqref{NEqdensity} for density. This discrepancy arises because the electron dynamics is governed by the full kinetic equation, including the fluctuation term in Eq.~\eqref{kineticequationelectrons}.For protons, by contrast, the kinetic equation~\eqref{kineticequationprotons} includes only the Vlasov term. However, unlike in the purely gravitational case described by Eq.\eqref{coarsegrainedstationarysolution}, the electrostatic field is now non-zero. This is clearly visible in the two bottom panels of Figure\ref{Compareforcesmix}, where we plot the individual force components acting on each species as functions of the spatial coordinate $\theta$. Specifically, in these panels we show:
\begin{itemize}
    \item in green: \begin{equation}\label{Elforce} F_{\alpha,1} = \mathrm{sign}(e_{\alpha}) , C \cdot Q \cdot \sin{\theta}, \end{equation}
    \item the external gravitational force in blue \begin{equation}\label{Extforce} F_{\alpha,2} = \tilde{g} , \sin\left(\frac{\theta}{2}\right), \end{equation}.
    \item the total force acting on a particle of species $\alpha$ in red: \begin{equation}\label{Totforce} F_{\alpha,3} = F_{\alpha,1} + F_{\alpha,2}. \end{equation}
\end{itemize}

As can be seen, the electrostatic force is of the same order of magnitude as the gravitational force for both electrons and protons, and thus plays a significant role in the system’s dynamics. Consequently, the density and temperature profiles obtained from simulations cannot match the analytical predictions of Eq.\eqref{NEqdensity} and Eq.\eqref{NEqtemperature}. However, an improved analytical expression can still be derived in this regime by incorporating the electrostatic potential into the single-particle Hamiltonians, as in Eq.~\eqref{Mean-field-Hamiltonians}. Replacing $\tilde{H}_\alpha$ with the full Hamiltonian $H_i$ in Eq.~\eqref{coarsegrainedstationarysolution}, we obtain the corrected coarse-grained distribution function
\begin{equation}\label{coarsegrainedstationarysolutionwithC}
    \tilde{f}_{i}(\theta,p)=\mathcal{N}_{i}\biggl(A\int_1^{+\infty}dT\frac{\gamma(T)}{T}e^{-\frac{H_{i}}{T}}+(1-A)e^{-H_{i}}\biggl),
\end{equation}
From this distribution, we compute the number density
\begin{equation}\label{NEqdensitywithC}
   \tilde{n}_{i}(\theta)=\frac{A\int_{1}^{+\infty}dT\frac{\gamma(T)}{\sqrt{T}}e^{-\frac{V}{T}}+(1-A)e^{-V}}{A\int_{1}^{+\infty}dT\frac{\gamma(T)}{\sqrt{T}}\int_{-\pi}^{\pi} d\theta e^{-\frac{V}{T}}+(1-A)\int_{-\pi}^{\pi} d\theta e^{-V}} \quad,
\end{equation}
and the kinetic temperature
\begin{equation}\label{NEqtemperaturewithC}
    \tilde{T}_{i}(\theta)=\frac{A\int_{1}^{+\infty}dT\gamma(T)\sqrt{T}e^{-\frac{V}{T}}+(1-A)e^{-V}}{A\int_{1}^{+\infty}dT\frac{\gamma(T)}{\sqrt{T}}e^{-\frac{V}{T}}+(1-A)e^{-V}} \quad.
\end{equation}
Here, the total potential $V$ acting on protons is
\begin{equation}
    V=2\tilde{g}\cos{\biggl(\frac{\theta}{2}\biggl)}+Cq(\cos{(\theta)}+1) \quad.
\end{equation}
The density (a decreasing function) and temperature (an increasing function) profiles given by Eqs.\eqref{NEqdensitywithC} and \eqref{NEqtemperaturewithC} are plotted in green in Fig. \ref{TempInvwithC2}. These show good agreement with the numerical results, demonstrating that the electrostatic field significantly influences proton dynamics. Furthermore, since the distribution functions in Eq.~\eqref{coarsegrainedstationarysolutionwithC} remain suprathermal and are governed by a mean-field Vlasov dynamics, the observed temperature inversion continues to be explained by the velocity filtration mechanism. Finally, even when the temperature is computed using the alternative definition in Eq.\eqref{coarsegrainedtempsuper}, a clear distinction between electron and proton behavior remains, as shown in Fig. \ref{TempInvwithC2}. Nevertheless, the proton temperature profile computed via Eq.\eqref{coarsegrainedtempsuper} coincides with that obtained using Eq.\eqref{coarsegrainedtemp}. This agreement confirms that proton fluctuations are negligible, and according to Eq.~\eqref{temperatureimbalance}, the two temperature definitions must converge. For electrons, by contrast, fluctuations are non-negligible, and the two definitions yield distinct temperature profiles. Moreover, since the dynamics for electrons are no longer governed by a purely Vlasov-type evolution, temperature inversion via gravitational filtering is no longer guaranteed across the entire spatial domain. This breakdown is clearly visible in the simulation results presented in the right-hand panels of Fig.~\ref{TempInvwithC2}.

\begin{figure}
    \centering
    \includegraphics[width=0.99\columnwidth]{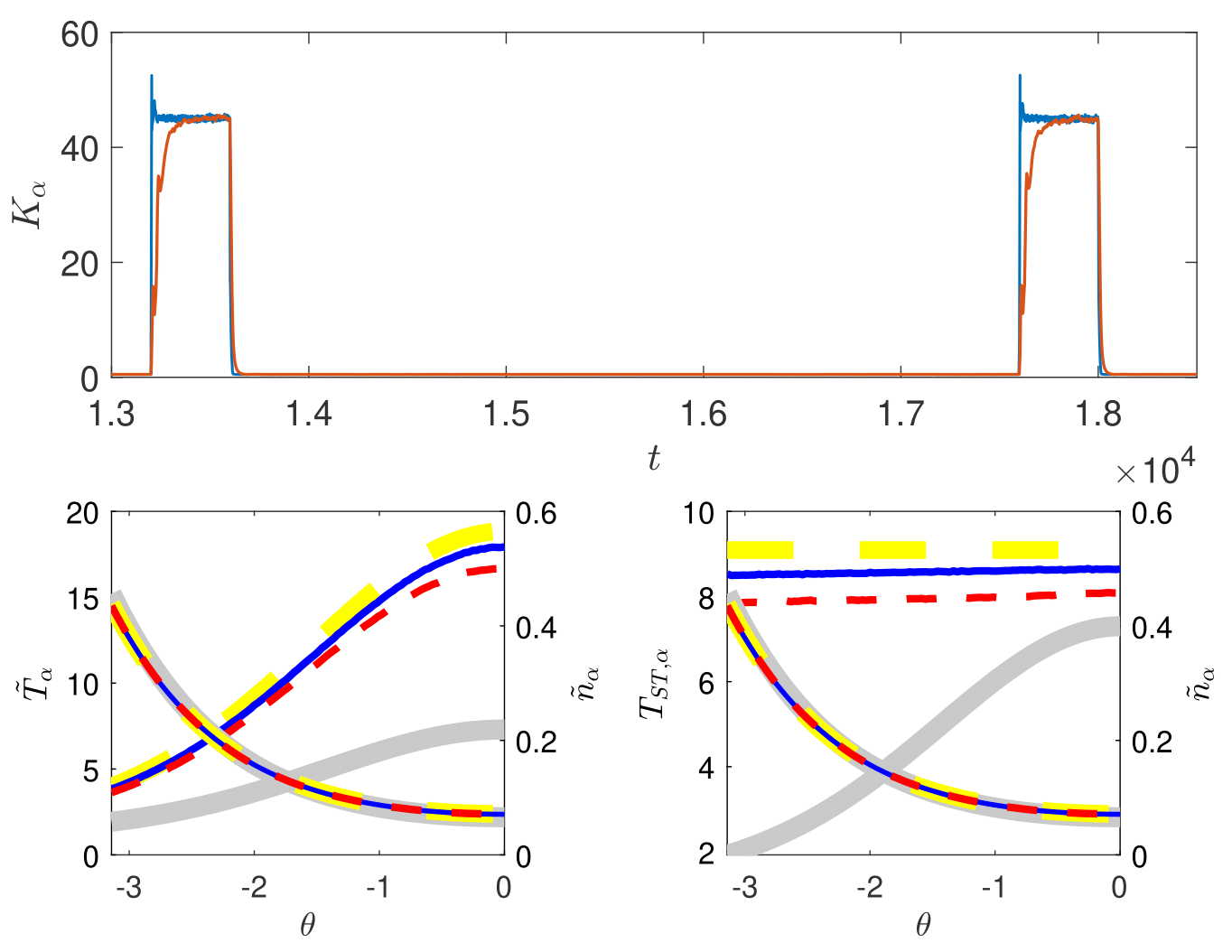}
    \caption{
    Top panel: Time series of the kinetic energies $K_{\alpha}$ for protons (orange) and electrons (blue), computed numerically using Eqs. \eqref{Kineticandstrat}. Bottom panels: Spatial profiles of the coarse-grained density and temperature for protons (dashed red) and electrons (solid blue), evaluated following the procedure outlined in Section~\ref{subsec:computationmoments}. In the left panel, temperature profiles are computed using Eq.\eqref{coarsegrainedtemp}; in the right panel, using Eq.\eqref{coarsegrainedtempsuper}. Theoretical predictions from Eq.\eqref{NEqdensity} (for density) and Eq.\eqref{NEqtemperature} (for temperature) are shown in grey. Dashed yellow curves correspond to the analytical profiles obtained by superposing multiple thermal configurations, as described in Section~\ref{subsec:Superpositiontheory}. Specifically, densities are computed using Eq.\eqref{superpositiondensities} (bottom panels), and temperatures using Eq.\eqref{superpositiontemperatures} (left) and Eq.~\eqref{superpositiontemperaturestrue} (right).
    }
    \label{Superpositionregime}
\end{figure}

\subsection{Superposition regimes}\label{subsec:superpositionregime}
By increasing the time scales $\tau$ and $t_w$ of the temperature fluctuations, the electrons will first reach a configuration of oscillation in time between various thermal equilibria, and then the protons will do so until the following regime is reached
\begin{equation}\label{supertempcondition}
    \tau, \langle t_w \rangle >>t_{R,p} >> t_{R,e}\quad.
\end{equation}
In this case, both the kinetic energies of protons and electrons oscillate around different thermal solutions, as illustrated in Fig. \ref{Superpositionregime}. In this regime both species are not in a stationary state, but since we are interested in observing the dynamics on a coarse-grained time scale given by Eq. \eqref{coarsegrainingcondition}, then the dynamics becomes stationary at such a coarse graining time scale.  All parameters have been fixed at the following numerical values: $\tau=400, t_w = 4000, \tilde{g}=1, M=100, C=400, \Delta T=90, N=2^{16}$ and integration time step $dt=4\cdot 10^{-3}$. As illustrated in Fig. \ref{Superpositionregime}, the coarse-grained numerical temperature profiles of both protons (in red) and electrons (in blue) calculated with both Eq. \eqref{coarsegrainedtemp} and Eq. \eqref{coarsegrainedtempsuper} exhibit significant discrepancies from the analytical solution (in gray) provided by Eq. \eqref{NEqtemperature} and they yield the profiles obtained from the superposition of disparate thermal solutions (in yellow), specifically calculated with Eq. \eqref{coarsegrainedtemp} for the left panel and with Eq. \eqref{coarsegrainedtempsuper} for the right panel. The same considerations are made in Eq. \eqref{coarsegrainedtempsuper} apply to the densities. In this instance, and in accordance with the theory outlined in Sec. \ref{subsec:Superpositiontheory}, the results are independent on the value of electrostatic coupling constant $C$. We also note that in this regime, a temperature inversion does or does not occur depending on how the coarse-grained temperature is calculated. In particular, if the temperature is calculated using the definition given by Eq. \eqref{coarsegrainedtempsuper}, then during the dynamics, the various temperatures are added together, resulting in a total temperature that is the superposition of the temperatures of many thermal states, as given by Eq. \eqref{superpositiontemperaturestrue}. Conversely, if the temperature is calculated using the definition given by Eq. \eqref{coarsegrainedtempsuper}, it corresponds to the temperature associated with the temporal coarse graining distribution function, which in this case is the superposition of many thermal distributions and is described by Eq. \eqref{VDFsuperpositiontemperatures}. As previously outlined in Section \ref{subsec:Superpositiontheory}, the distribution functions in question exhibit suprathermal tails. Given that it is determined at each point by the conservation of energy of a single particle, $\tilde{H}_{\alpha}$, gravitational filtering leads to the emergence of a monotonically increasing temperature profile that is anti-correlated to the density profile.

\section{Summary and perspectives}\label{sec5}

In this work, we have extended the analysis of the dynamics of a stratified, collisionless plasma gravitationally confined in a loop and in thermal contact with a fluctuating boundary beyond the previously studied regime where the fluctuation time scales are much smaller than the relaxation times of electrons and protons. Building upon previous investigations by \citealt{barbieri2023temperature,Barbieri2024b}, we have developed a generalized temporal coarse-graining framework that remains valid when the characteristic time scales of the temperature fluctuations are no longer much shorter than the electron crossing time. By deriving a set of coarse-grained kinetic equations, we have identified a novel contribution, arising directly from the coarse-graining procedure that affects the stationary profiles of the plasma. Our numerical results show that when the time scale of temperature fluctuations becomes comparable with the electron crossing time, this additional term leads to a species-dependent separation in the temperature and density profiles. In this regime, the stationary state of the system is no longer described by the effective distribution function introduced in \citealt{Barbieri2024b} and gravitational filtering is no longer sufficient to guarantee temperature inversion. This effect is interpreted through the emergence of electrostatic fluctuations that generate a correction to the standard Vlasov dynamics. We have also investigated the opposite limit, in which the time scales of temperature fluctuations are much longer than the crossing times of both species. In this regime, the system evolves through a sequence of thermal states, and its long-term behaviour can be described by a coarse-grained distribution that is a superposition of thermal solutions corresponding to different boundary temperatures.

We have also introduced two definitions of temperature at the temporal coarse-graining level. The first is defined as the ratio between the coarse-grained pressure and the coarse-grained density. The second is obtained by computing the instantaneous kinetic temperature at each time step and then averaging over time. The first definition assumes that measurements of the moments can be performed with a resolution much shorter than the coarse-graining time scale, thereby allowing access to the system’s microscopic dynamics. In contrast, the second definition is appropriate when the system dynamics below the coarse-graining scale $\tilde{t}$ are not accessible. In this regime, the behaviour of the two species is described by the coarse-grained distribution functions $\tilde{f}_{\alpha}$, and the kinetic temperature is computed from these via the standard kinetic definition. The two definitions yield the same result only when fluctuations are negligible, that is, in the stationary regime. In other regimes, particularly in the superposition regime, a temperature inversion persists if the temperature is computed as the ratio of coarse-grained pressure to coarse-grained density. However, if the coarse-grained temperature is instead defined as the average of the instantaneous kinetic temperatures, the resulting profile becomes isothermal. Such an inversion is expected to be observed when the integration time of the measurement significantly exceeds the relaxation time.

As an initial avenue for future investigation, it would be worthwhile to extend the HMF approximation for the electrostatic field by incorporating multiple Fourier modes in the expansion. Similarly, the temporal coarse-graining procedure could be generalized to account for a multi-mode electrostatic interaction. This extension would allow us to address several key questions: Does the overall dynamics depend on the specific form of the electrostatic interaction? And, more importantly, does the plasma always reach the stationary regime described in Sec. \ref{subsec:stationarystateregime}? This latter question is intriguing not only from a theoretical perspective but also for its implications in solar physics. As highlighted earlier, our model applied to the solar atmosphere (see, e.g.,\citealt{barbieri2023temperature}) demonstrates that rapid, intense, intermittent, and short-lived heating events can drive the plasma toward a stationary configuration characterized by anti-correlated density and temperature profiles similar to those observed in the Sun. An affirmative answer to this question would provide strong evidence of the robustness of our results. Another promising extension is the inclusion of a magnetic field along the loop axis in the plasma model. Such an addition is of interest not only for modeling solar coronal loops but also in the context of fusion plasmas confined in Tokamak devices (see, e.g., \citealt{goedbloed_keppens_poedts_2010,2018CoPP...58..457C}). Introducing an axial magnetic field, which would naturally decrease in strength from the base to the top of the loop due to expansion with altitude, would impose the additional constraint of magnetic moment conservation on the particle motion. This conservation would tend to increase the parallel temperature while decreasing the perpendicular temperature, thereby enhancing gravitational filtering along the magnetic field and diminishing it in the orthogonal direction. With this extra ingredient, several new questions arise: How does the temperature inversion change in the steady-state configuration? For what values of the loop expansion rate does the perpendicular temperature still increase monotonically with height? And how does the overall plasma dynamic picture change?

Finally, it is important to note that throughout this discussion the plasma has been treated as collisionless. The framework presented in this paper may be affected by the introduction of collisions in the first two regimes, but not in the superposition regime described in Sec. \ref{subsec:superpositionregime}, since in that case the system oscillates between various thermal configurations that are also solutions of the Boltzmann equation. In the context of solar coronal applications, the introduction of Coulomb collisions becomes particularly relevant. The main questions to be addressed are: Beyond the superposition regime, how does plasma dynamics change with the inclusion of collisions? Is it always possible to identify a stationary state regime comparable to that described in Sec. \ref{subsec:stationarystateregime}? If so, how do the properties of this regime evolve under collisional effects? Under what conditions do temperature inversion and non-thermal distributions persist in the presence of Coulomb collisions?

\begin{acknowledgments}\textbf{Acknowledgments.} The authors wish to thank the anonymous reviewers for their insightful comments, which improved the presentation of the work. L.B. wishes to thank Arnaud Zaslavsky for valuable discussions.
\end{acknowledgments}

\begin{acknowledgments}\textbf{Funding.}
We acknowledge the Fondazione CR Firenze under the projects \textit{HIPERCRHEL}.
This research was partially funded by the European Union - Next Generation EU - National Recovery and Resilience Plan (NRRP) - M4C2 Investment 1.4 - Research Programme  CN00000013 "National Centre for HPC, Big Data and Quantum Computing" - CUP B83C22002830001 and by the European Union - Next Generation EU - National Recovery and Resilience Plan (NRRP)- M4C2 Investment 1.1- PRIN 2022 (D.D. 104 del 2/2/2022) - Project `` Modeling Interplanetary Coronal Mass Ejections'', MUR code 31. 2022M5TKR2,  CUP B53D23004860006.
Views and opinions expressed are however those of the author(s) only and do not necessarily reflect those of the European Union or the European Commission. Neither the European Union nor the European Commission can be held responsible for them.  L.B. wants to thank
the Sorbonne Université in the framework of the Initiative Physique des Infinis for financial support.
\end{acknowledgments}

\begin{acknowledgments}\textbf{Declaration of interests.}
The authors report no conflict of interest.
\end{acknowledgments}

\bibliographystyle{jpp}
\bibliography{jpp-instructions}
\end{document}